\documentclass{article}

\usepackage[english]{babel}

\usepackage[a4paper, top=2cm, bottom=2cm, left=3cm, right=1cm, marginparwidth=0cm, nohead]{geometry}

\usepackage[fontsize=14pt]{scrextend}
\usepackage{setspace}
\usepackage{indentfirst}
\usepackage{amsfonts}
\usepackage{amsmath}
\usepackage{amssymb}
\usepackage{amsthm}
\usepackage{graphicx}
\usepackage{float}
\usepackage[colorlinks=true, allcolors=blue, linktocpage=true]{hyperref}
\usepackage{braket}
\usepackage{centernot}
\usepackage[shortlabels]{enumitem}
\usepackage{longtable}
\usepackage{units}
\usepackage{caption}
\usepackage{listings}
\usepackage[table, svgnames]{xcolor}
\usepackage{secdot}
\usepackage{sectsty}
\usepackage{multirow}
\usepackage{multicol}
\usepackage{algorithm}
\usepackage[noEnd=false, indLines=true, commentColor=black]{algpseudocodex}
\usepackage{dirtytalk}
\usepackage[figurename=Fig., labelsep=period]{caption}

\addto\captionsenglish{}

\addto\captionsenglish{}

\addto\captionsenglish{}

\addto\captionsenglish{}

\addto\captionsenglish{}

\newcommand\myhash{\scalebox{0.7}{\raisebox{0.35ex}{\#}}}

\makeatletter
\def\ps@plain{%
      \let\@oddhead\@empty
      \let\@evenhead\@empty
      \def\@oddfoot{\normalfont\hfil\thepage}%
      \def\@evenfoot{\normalfont\hfil\thepage}}
\makeatother
\definecolor{backcolour}{rgb}{1.0,1.0,1.0}
\definecolor{codenumber}{rgb}{0.14,0.47,0.69}
\definecolor{codecomment}{rgb}{0,0.50,0}
\definecolor{codekeyword}{rgb}{0,0,1}
\definecolor{codestring}{rgb}{0.72,0.08,0.08}

\lstdefinestyle{mystyle}{
  backgroundcolor=\color{backcolour}, 
  commentstyle=\color{codecomment},
  keywordstyle=\color{codekeyword},
  numberstyle=\tiny\color{codenumber},
  stringstyle=\color{codestring},
  basicstyle=\ttfamily\small,
  breakatwhitespace=false,         
  breaklines=true,                 
  captionpos=b,                    
  keepspaces=true,                 
  numbers=left,                    
  numbersep=5pt,                  
  showspaces=false,                
  showstringspaces=false,
  showtabs=false,                  
  tabsize=2
}

\sectionfont{\fontsize{14}{14}\selectfont}
\subsectionfont{\fontsize{14}{14}\selectfont}

\begin{document}
\setstretch{1.3}
\thispagestyle{empty}


\begin{center}

\vspace{2cm}

\textbf{National Research University Higher School of Economics}

\textbf{Faculty of Computer Science}

\textbf{Programme ‘Master of Data Science’}

\vspace{3cm}

\textbf{MASTER’S THESIS}

\vspace{3cm}

\Large
\textbf{Electricity Spot Prices Forecasting Using\\ Stochastic Volatility Models}

\vspace{4cm}

\normalsize
\textbf{Student} \ \ \ \ \ \ \ \ \ \ \ \ \ \ \ \ \ \ \ \ \ \ \ \ \ \ \ \ \ \ \ \ \ Batyrov Andrei Renatovich

\vspace{1cm}

\textbf{Supervisor} \ \ \ \ \ \ \ \ \ \ \ \ \ \ \ \ \ \ \ \ \ Kasianova Ksenia Alekseevna

\vfill
         
\textbf{Moscow, 2024}            
\end{center}

\newpage

\begin{center}
\textbf{ABSTRACT}
\end{center}

There are several approaches to modeling and forecasting time series as applied to prices of commodities and financial assets. One of the approaches is to model the price as a non-stationary time series process with heteroscedastic volatility (variance of price).

The goal of the research is to generate probabilistic forecasts of day-ahead electricity prices in a spot marker employing stochastic volatility models. A typical stochastic volatility model -- that treats the volatility as a latent stochastic process in discrete time -- is explored first. Then the research focuses on enriching the baseline model by introducing several exogenous regressors.

A better fitting model -- as compared to the baseline model -- is derived as a result of the research. Out-of-sample forecasts confirm the applicability and robustness of the enriched model. This model may be used in financial derivative instruments for hedging the risk associated with electricity trading.


\newpage
\tableofcontents
\newpage
\listoffigures
\newpage
\listoftables
\newpage
\listofalgorithms
\newpage

\section{INTRODUCTION}

Today electricity is traded in markets around the world using spot and derivative contracts. Electricity demand depends on individual and business activities, as well as weather and other factors, which may create various seasonality profiles: peak and off-peak hours, working days and holidays, etc.

Electricity is non-storable which makes it a unique commodity and leads to its price having high volatility with possible sudden spikes (shocks), which in turn leads to the need of hedging the risk associated with trading this commodity. This is why predicting the price at least one day ahead is so important.

In the studied scenario, the electricity spot market is divided into two price zones subject to economic and climate factors:
\begin{enumerate}[noitemsep]
\item European market;
\item Siberian market.
\end{enumerate}
Each market has its own consumer consumption and price profiles. In this research we will study hourly profiles, that is we will build and examine hourly (high frequency) models for:
\begin{enumerate}[noitemsep]
\item Peak hour;
\item Off-peak hour. 
\end{enumerate}

\subsection{Approaches to Forecasting}

Electricity price forecasting uses mathematical, statistical and machine learning models to predict electricity prices in the future. As of today, several approaches and models have been proposed and tested \cite{8}:
\begin{itemize}[noitemsep]
\item Multi-agent (multi-agent simulation, equilibrium, game theoretic);
\item Fundamental (structural);
\item Reduced-form (quantitative, stochastic);
\item \textit{Statistical (econometric, technical analysis)};
\item Computational intelligence (artificial intelligence-based, non-parametric, non-linear statistical);
\item Hybrid solutions.
\end{itemize}

Statistical approach includes the following models \cite{8}:
\begin{itemize}[noitemsep]
\item Similar-day and exponential smoothing;
\item Regression (AR, ARMA, etc.);
\item \textit{Heteroscedastic (non-constant volatility)}.
\end{itemize}

Among heteroscedastic models there are:
\begin{itemize}[noitemsep]
\item (Generalized) Autoregressive Conditional Heteroscedasticity ((G)ARCH);
\item Stochastic Volatility (SV).
\end{itemize}

In (G)ARCH models, the variance of the time series is represented by an autoregressive (ARCH) or autoregressive with moving average (GARCH) process \cite[p. 1055]{8}, which means that the volatility is deterministic at time $t$. On the contrary, in SV models, the volatility is modeled as a stochastic, that is random, process \cite[p. 361]{1}.

A comprehensive comparison of GARCH vs SV models is studied in \cite{7} with Bayesian estimation of model parameters. The authors claim that \say{... stochastic volatility models almost always outperform their GARCH counterparts, suggesting that stochastic volatility models might provide a better alternative to the more conventional GARCH models.}

This research is devoted to building and enriching stochastic volatility models as applied to electricity prices forecasting in a spot market.

\subsection{Acronyms}

The following is a list of acronyms used in this thesis paper.

\begin{longtable}[l]{rl}
\textit{ADF} & Augmented Dickey-Fuller (test) \\
\textit{AR} & AutoRegressive (model) \\
\textit{CI} & Confidence Interval \\
\textit{ICE} & Individual Conditional Expectation \\
\textit{IDE} & Integrated Development Environment \\
\textit{MAE} & Mean Absolute Error \\
\textit{ML} & Machine Learning \\
\textit{MSE} & Mean Squared Error \\
\textit{MWU} & Mann-Whitney U (test) \\
\textit{OOP} & Object-Oriented Programming \\
\textit{PACF} & Partial Autocorrelation Function \\
\textit{PD} & Partial Dependence \\
\textit{PPD} & Posterior Predictive Distribution \\
\textit{RV} & Random Variable \\
\textit{RMSE} & Root Mean Squared Error \\
\textit{SV} & Stochastic Volatility \\
\textit{SV X} & Stochastic Volatility Exogenous (model) \\
\end{longtable}

\newpage

\section{MODEL IMPLEMENTATION}

\subsection{Bayesian Inference}

Model parameters estimation can be a challenging task. By far not always parameters can be expressed in closed-form solutions. One of the approaches to estimate (learn) model parameters in ML is Bayesian inference which treats model parameters as RV, that is each parameter is described by its distribution with some density rather than by a single value \cite[p. 139]{15}. The general framework of Bayesian inference is the following:
\begin{enumerate}[noitemsep]
\item Choose a prior distribution for the parameter $p(\theta)$. Prior distribution is a distribution before the train data is observed. In our study the train data is the price and may also include exogenous regressors, altogether denoted as $y$ here;
\item Given the train data -- the distribution of the observed data conditional on its parameters, i.e. $p(y | \theta )$, also known as the likelihood function $\operatorname {L} (\theta | y ) = p(y | \theta )$, --
\item Generate the posterior distribution of the parameter with density
\begin{equation} \label{eq:1}
p(\theta | y) = \frac{p(y | \theta) p (\theta)}{\int{p(y | \theta) p (\theta) d \theta}}.
\end{equation}
\item To generate predictions for new unseen data $\tilde{y}$, compute the PPD with density
\begin{equation} \label{eq:2}
p(\tilde{y} | y) = \int{p(\tilde{y} | \theta) p (\theta | y) d \theta}.
\end{equation}
Thus, instead of a single point as a prediction, a distribution is generated -- same concept as with the model parameter distribution.
\end{enumerate}

\subsection{Development Environment}

Python \cite{22} was chosen as the programming language to perform all accompanying mathematical computations to support this research. An interactive computations approach was chosen through the use of Jupyter Notebooks \cite{23} which were developed within an IDE Visual Studio Code \cite{24}. The code for all computations was developed, tested and run on Linux (Ubuntu) in a Docker container \cite{26} running on a Windows machine.

Along with Python -- to ensure correctness and robustness of the computations results and to facilitate the speed of development and debugging -- industry- and community-proven tools and libraries were chosen and used:
\begin{itemize}[noitemsep]
\item Stan \cite{4} with Python interface PyStan \cite{27}: Bayesian inference;
\item scikit-learn \cite{11}: OOP approach for model building and cross-validation, clustering, metrics;
\item Pandas \cite{19}: data manipulation, matrix and vector operations, descriptive statistics;
\item NumPy \cite{18}: polynomial fitting, correlation;
\item SciPy \cite{17}: hypothesis testing;
\item statsmodels \cite{16}: hypothesis testing, PACF;
\item Matplotlib \cite{20}: plotting;
\item seaborn \cite{21}: plotting.
\end{itemize}

A Python module was developed for implementing a custom class \\ \texttt{StanModel} which inherits from the scikit-learn's \texttt{BaseEstimator} and \\ \texttt{RegressorMixin} classes. This allows to overload the scikit-learn's standard methods, such as \texttt{fit()}, \texttt{predict()}, etc. To correctly split the times series data into train-test folds, the \texttt{TimeSeriesSplit} class was used. To ensure reproducibility of the results, the random seed was fixed.

All computation code supporting this paper can be found at the author's GitHub repository \cite{10}.

\subsection{Modeling with Stan}

For model parameters estimation we will use the Stan platform for statistical modeling and high-performance statistical computation. It uses its own proprietary probabilistic language and can do Bayesian statistical inference and prediction. Other notable probabilistic programming library is PyMC. The main blocks of a Stan program are the following:
\begin{itemize}[noitemsep]
\item data: declaration of variables that are read in as external data;
\item parameters: declaration of model parameters that will be learned during  inference;
\item model: model definition and generation of model parameters posterior distributions -- fitting the model;
\item generated quantities: generation of PPD -- predicting for new data with the fitted model.
\end{itemize}

Our custom class \texttt{StanModel} implements two main methods: \texttt{fit()} and \texttt{predict()}, which are wrappers around PyStan's \texttt{sample()} and \\ \texttt{fixed\_param()} methods and provide scikit-learn-compatible interfaces. The \texttt{sample()} method is used for generating model parameters posterior distributions (fit) and \texttt{fixed\_param()} method is used for generating PPD (predict).

Generating posterior distributions may be time consuming. Taking into account the need to fit and predict many models during building, cross-validation and forecasting, we would like to optimize the running time of Stan code. A possible approach for generating PPD for predictions is to use only the expected values of the estimated parameters, instead of their full posterior distributions. Experiments show that the posterior distributions of model parameters are almost always close to normal (see Figures \ref{fig:sv_base_4} and \ref{fig:sv_x_5}) which are symmetric and has one mode only. This approach might not be acceptable for severely non-normal distributions with long tails or several modes, though. The density of a now constant parameter is a Dirac delta function shifted by the expected value of the parameter. Thus, instead of \ref{eq:2}, the density of PPD will be computed as
\begin{equation} \label{eq:3}
p(\tilde{y} | y) \approx \int{p(\tilde{y} | \theta) \delta(\theta - \mathbb{E} [\theta | y]) d \theta}.
\end{equation}

Practically, this means using the mean values of the estimated parameters when drawing samples from PDD with Stan. When examining predictions, we will use CI to compensate for the "loss" of full parameters distributions (estimation uncertainty). Stan example code for a basic SV model can be found at \cite{4}, \cite{5}. Complete Stan code for SV Baseline and SV X models developed in this research can be found in the Appendix.

\newpage

\section{SV BASELINE MODEL}

There is a common practice in ML tasks to begin the research with the so-called baseline model. The baseline model is used as a foundation for more complex models to build upon. A baseline model can be simple with low variance and/or high bias and not necessarily be a good estimator, thus enabling it to be used as a reference model to estimate the quality and improvement, if any, of all other derived models.

A typical SV model can be described with the following parameters \cite[p.361]{1}, \cite{4}:
\begin{itemize}[noitemsep]
\item $\mu$, mean log volatility;
\item $\phi$, persistence of volatility;
\item $\sigma$, white noise shock scale;
\item $h_{t}$, latent log volatility at time $t$.
\end{itemize}
The variable $\epsilon_{t}$ represents the white-noise shock (i.e., multiplicative error) on the price at time $t$, whereas $\delta_{t}$ represents the shock on volatility at time $t$:
$\epsilon_{t} \sim \mathcal{N}(0, 1)$; $\delta_{t} \sim \mathcal{N}(0, 1)$.

Then the price at time $t$ can be described as
\begin{equation} \label{eq:4}
y_{t} = e^{h_{t}/2} \epsilon_{t},
\end{equation}
where $\displaystyle h_{t} = \mu + \phi (h_{t-1} - \mu) + \delta_{t} \sigma$;
$\displaystyle h_{1} \sim \mathcal{N} \left( \mu, \frac{\sigma}{\sqrt{1 - \phi^2}} \right)$.

Rearranging the equations above yields the following final model for the price:
\begin{equation} \label{eq:5}
y_{t} \sim \mathcal{N} \left ( 0, e^{h_{t} / 2} \right ).
\end{equation}
The time point $t$ is subject to desired market frequency. In our research we examine hourly profiles distributed over 24 hours (days), that is each point in time represents exactly one consumer price during the whole given hour on a given day.

\subsection{Consumer Prices}

As explained in the Introduction, there are two price zones: 1 (European) and 2 (Siberian) in the day-ahead spot electricity markets. We will use the following datasets in our research:
\begin{enumerate}[noitemsep]
\item For modeling, we will load the hourly data for the price zone 1 for the period of 01.05.2023 -- 30.04.2024 (one year back from now) for the peak hour.
\item For model cross-validation, we will load and examine both price zones for both peak and off-peak hours with cross-validation over a sliding window for the period of 23.06.2014 -- 30.04.2024 (approx. ten years back from now).
\item For forecasting, we will generate predictions for both price zones for both peak and off-peak hours for the period of 01.05.2024 -- 07.05.2024 (one week ahead).
\end{enumerate}
The Daily Indices and Volumes of The Day-ahead Market data is available on the Administrator of Trading System (ATS) \cite{3}, starting from Aug 8th, 2013 to present. The unit of prices is RUB/MWh.

\begin{figure}
\centering
\includegraphics[scale=0.75]{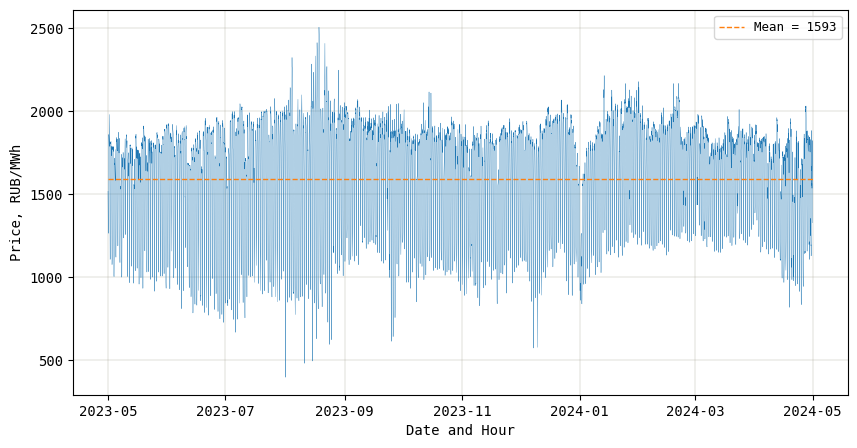}
\caption{\label{fig:sv_base_1}Consumer prices}
\end{figure}

A plot of all consumer prices for price zone 1 (modeling dataset) is shown in the Figure \ref{fig:sv_base_1}. The market time frame is one hour.

\begin{figure}
\centering
\includegraphics[scale=0.8]{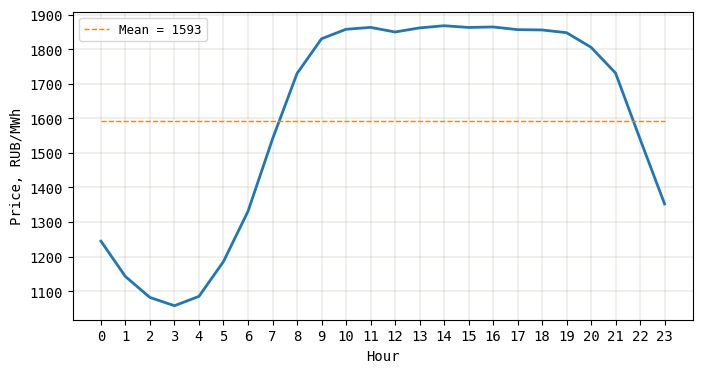}
\caption{\label{fig:sv_base_2}Consumer prices mean hourly profile}
\end{figure}

As explained in the Introduction, we are modeling the price during each hour. The plot of mean prices for each hour is shown in the Figure \ref{fig:sv_base_2}.

\begin{figure}
\centering
\includegraphics[scale=0.55]{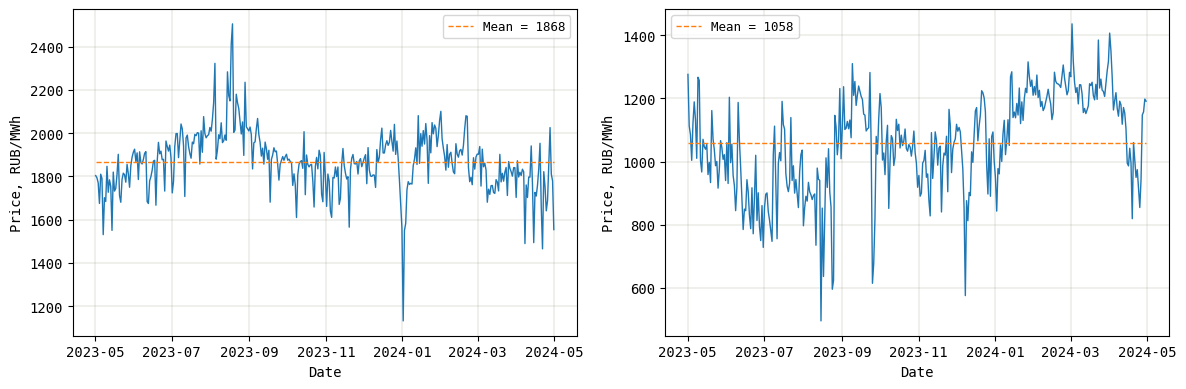}
\caption{\label{fig:sv_base_3}Trace plots for peak hour \myhash 14 (left) and off-peak hour \myhash 3 (right)}
\end{figure}

The hourly profile, that is the change of the price during the 24 hours (one day) is clearly seen. We will choose one peak hour \myhash 14 and one off-peak hour \myhash 3. The consumer prices for both peak and off-peak hours over the whole modeling time frame are shown in the Figure \ref{fig:sv_base_3}.

\subsection{Test For Stationarity} \label{subsec_test_stationarity}

The plots in the Figure \ref{fig:sv_base_3} suggest that the price is not a stationary process. We will check this hypothesis using the ADF test \cite{6}, \cite[p. 2]{13}. This test tests the null hypothesis that there is a unit root in the time series. A linear stochastic process has a unit root if the process's characteristic equation has a root equal to 1, hence the name. Such a process is non-stationary. The alternative hypothesis is usually stationarity.

$\alpha = 0.05$

$\mathcal{H}_{0}$: Unit root (Non-stationary)

$\mathcal{H}_{1}$: Stationary

The results of the ADF tests are shown in the Table \ref{tab:sv_base_1}.

\begin{longtable}{r||r|r|l}
\hline
Hour & Statistic & $p$-value & Result at $\alpha$ \\
\hline \hline
Peak \myhash 14 & -2.56 & 0.10 & Non-stationary \\
Off-peak \myhash 3 & -2.71 & 0.07 & Non-stationary \\
\hline
\caption{\label{tab:sv_base_1}ADF tests results}
\end{longtable}

We cannot reject the null hypothesis at the significance level of 0.05. The consumer price is a non-stationary process, which means that the variance of price (volatility) is an RV with its own moments and can thus be modeled as a stochastic process itself.

\subsection{Modeling}

The mean of $y_{t}$ is modeled as $0$ in the original model \ref{eq:5}, since this model was derived for returns on holding an asset, that is the difference between two consecutive prices, as per the market time frame. In our research we will focus on studying the price itself but not the returns, which means that we have to change the model's mean from $0$ to our data real mean, which yields the following final SV Baseline model:
\begin{equation} \label{eq:6}
y_{t} \sim \mathcal{N} \left (\bar{y}, e^{h_{t} / 2} \right ),
\end{equation}
where $\bar{y}$ is the mean price seen during estimation (learning) of the model's parameters. The consumer price at time $t$ is a sample drawn from the normal distribution with the mean value being the train price mean (constant) and the variance being the exponent of half of the train log volatility, again at time $t$ (stochastic).

We have shown that the price is a non-stationary process (see \ref{subsec_test_stationarity}), which means that both the mean and variance are not constant over time. Strictly speaking, we should not assume the constant mean $\bar{y}$ in \ref{eq:6}. However, since this is a baseline model, we will keep the assumption of constant mean, yet stochastic volatility at the same time. We will address this issue with the mean later while developing an extended model.

\begin{figure}
\centering
\includegraphics[scale=0.75]{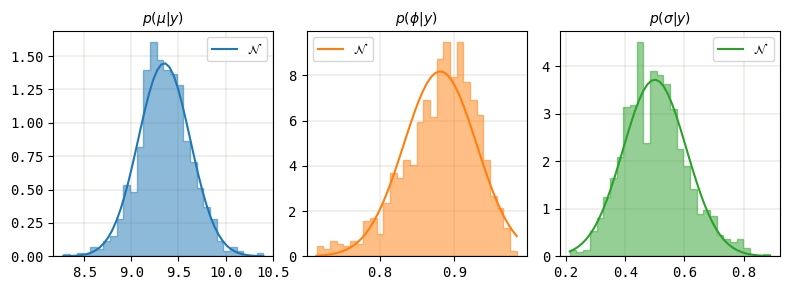}
\caption{\label{fig:sv_base_4}SV Baseline: Posterior distributions of estimated parameters}
\end{figure}

After running the Stan code, we have obtained the posterior distributions of all SV Baseline model's parameters -- see Figure \ref{fig:sv_base_4}.

\begin{figure}
\centering
\includegraphics[scale=0.8]{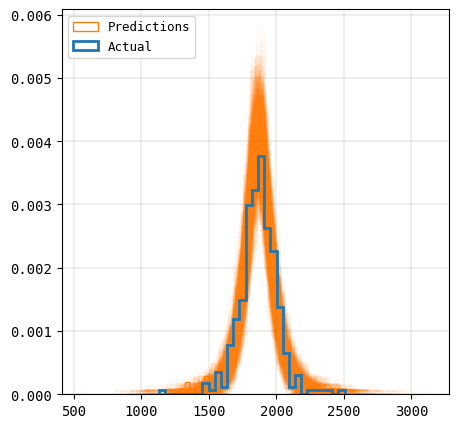}
\caption{\label{fig:sv_base_11}SV Baseline: 1,000 predicted and actual price distributions}
\end{figure}

\begin{figure}
\centering
\includegraphics[scale=0.75]{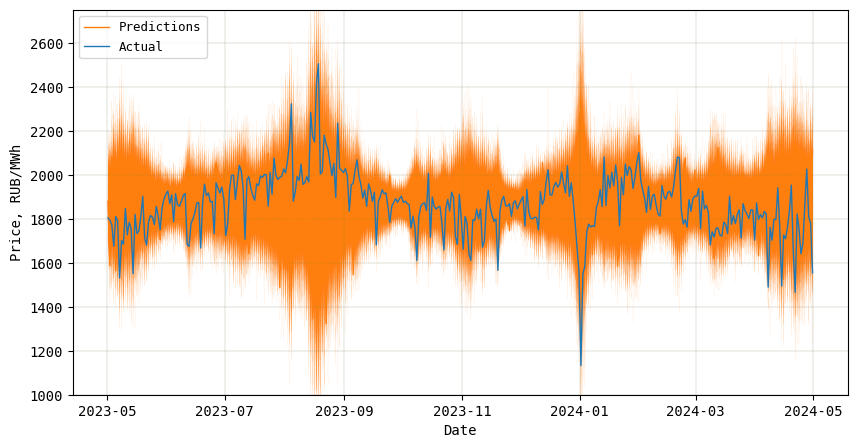}
\caption{\label{fig:sv_base_5}SV Baseline: Actual price and 1,000 predictions}
\end{figure}

\begin{figure}[hb]
\centering
\includegraphics[scale=0.75]{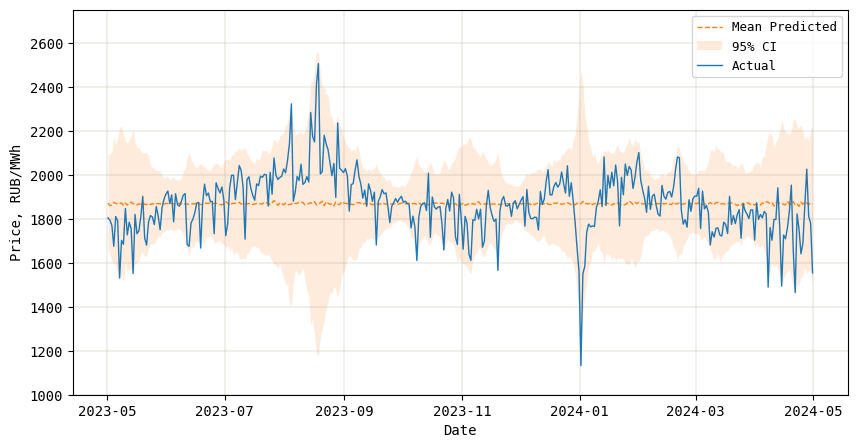}
\caption{\label{fig:sv_base_6}SV Baseline: 95\% CI for 1,000 predictions}
\end{figure}

We can now use the estimated model's parameters to compute PDD and generate in-sample predictions for the modeling time frame. 1,000 density plots sampled from the PDD of the consumer price are shown in the Figure \ref{fig:sv_base_11}. Observe, that the actual distribution (density) of the consumer prices is well within the predicted density plots, which confirms the correctness of modeling. 1,000 trace plots sampled from the PDD of the consumer prices are shown in the Figure \ref{fig:sv_base_5}. The 95\% CI for all 1,000 predictions is shown in the Figure \ref{fig:sv_base_6}.

\begin{figure}
\centering
\includegraphics[scale=0.70]{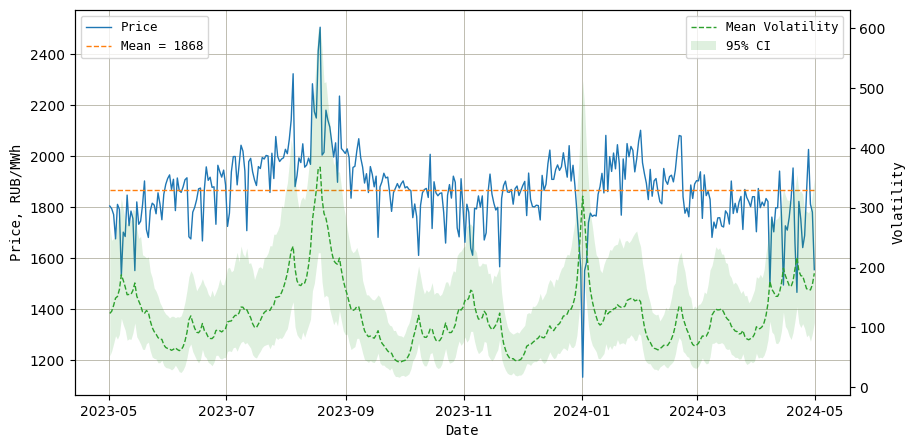}
\caption{\label{fig:sv_base_7}SV Baseline: Consumer prices and learned volatility}
\end{figure}

The model has learned the latent stochastic volatility process quite correctly. The learned volatility $e^{h_{t} / 2}$ (both mean and 95\% CI) is shown in the Figure \ref{fig:sv_base_7}.

\begin{figure}
\centering
\includegraphics[scale=0.8]{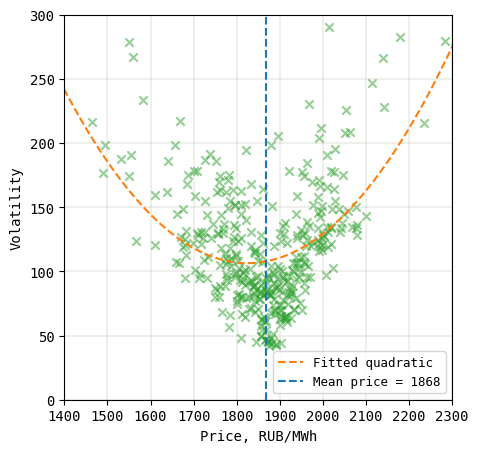}
\caption{\label{fig:sv_base_8}SV Baseline: Learned volatility vs Consumer price}
\end{figure}

To understand the behavior of the volatility we will visually check the correlation between the price and volatility -- see Figure \ref{fig:sv_base_8}. We can clearly see that volatility \textit{does} depend on consumer price, having a rather V-shaped dependency: the volatility tends to increase for the prices higher and lower than the mean price, while it's minimal for the prices around the mean.

\subsection{Goodness-of-fit}

To evaluate the quality of the model we will compute the following metrics \cite[p. 28, 437]{12}, \cite[p. 1039]{8}:
\begin{equation} \label{eq:7}
MAE(y, \hat{y}) = \frac{1}{n} \sum_{i=1}^{n} |y_{i} - \hat{y_{i}}|,
\end{equation}

\begin{equation} \label{eq:8}
RMSE(y, \hat{y}) = \sqrt{MSE(y, \hat{y})} = \sqrt{\frac{1}{n} \sum_{i=1}^{n} (y_{i} - \hat{y_{i}})^2},
\end{equation}
where $n$ is the number of target points, $y_{i}$ are the true target values, $\hat{y_{i}}$ are the model's predictions.

We will apply the metrics \ref{eq:7} and \ref{eq:8} to the consumer price. Note that both MAE and RMSE are point metrics. We are drawing 1,000 samples from PDD which are then averaged for each time $t$, and these averaged consumer prices are used in the metrics.

The metrics for the in-sample predictions made by the SV Baseline model are shown in the Table \ref{tab:sv_base_2}.

\begin{longtable}{l||r}
\hline
Metric & SV Baseline\\
\hline \hline
MAE & 99.18 \\
RMSE & 137.04 \\
\hline
\caption{\label{tab:sv_base_2}SV Baseline: MAE and RMSE for predicted consumer prices}
\end{longtable}

\begin{figure}
\centering
\includegraphics[scale=0.8]{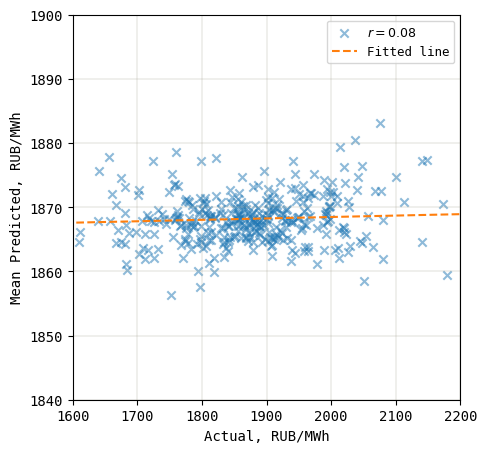}
\caption{\label{fig:sv_base_9}SV Baseline: Predicted vs Actual price}
\end{figure}

We also want to check the correlation between the predicted mean price and actual price -- see Figure \ref{fig:sv_base_9}.

Though the model was able to learn heteroscedasticity of the volatility, the correlation between the mean predictions and actual prices is rather weak.

\begin{figure}[ht]
\centering
\includegraphics[scale=0.55]{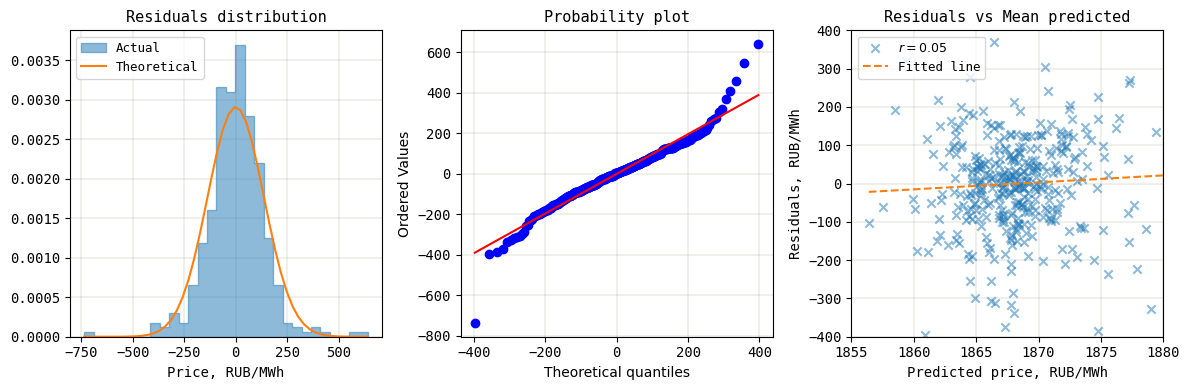}
\caption{\label{fig:sv_base_10}SV Baseline: Residuals plots}
\end{figure}

We should also check the distribution of residuals, probability plot, and correlation between residuals and mean predicted prices -- see Figure \ref{fig:sv_base_10}. Residuals look to be distributed more or less normally with several outliers. Also, residuals are almost not correlated with predictions (homoscedastic).

We will \textit{accept} the SV Baseline model.

\newpage

\section{SV EXOGENOUS MODEL}

Having built our baseline model, we want to derive a better performing model. One of the approaches is adding independent variables that should help explain the price change over time. We will call such variables as exogenous regressors.

The idea of enriching a model by introducing exogenous factors was studied in \cite{2}. The authors introduce the logarithm of the hourly air temperature at time $t$, indicator variables for days of the week, and the minimum of the previous day's 24 hourly log prices as the exogenous regressors.

We will now examine similar regressors to understand if they can enrich our model to provide better quality.

\subsection{Air Temperature}

\begin{figure}
\centering
\includegraphics[scale=0.75]{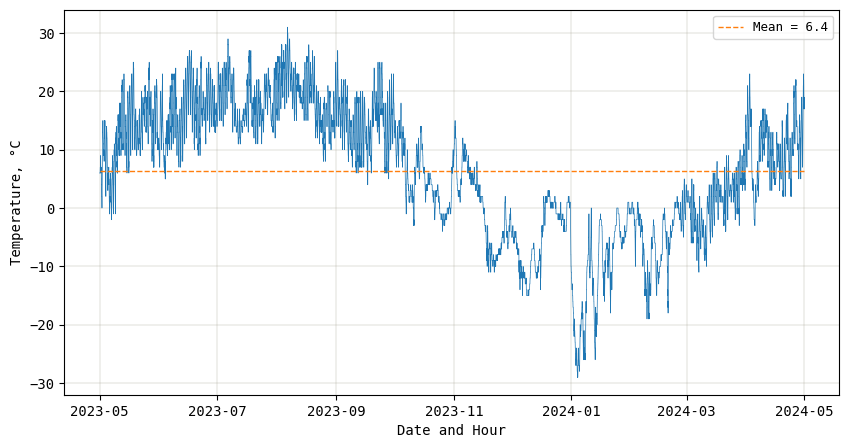}
\caption{\label{fig:sv_x_1}Air temperature in Moscow}
\end{figure}

We will start with the air temperature variable. It may well be expected that demand for electricity, and thus its price, can be dependent on air temperature. Heating is required in cold season (winter) and cooling is required in hot season (summer).
The air temperature and other weather data can be found at \cite{14}. Archives of hourly data for different countries and places are available. The hourly air temperature profile in Moscow -- price zone 1 over the modeling time frame -- is shown in the Figure \ref{fig:sv_x_1}. Each point on the plot represents the air temperature for one whole hour. We are studying the same one-hour time frame as we did for the consumer prices.

\begin{figure}
\centering
\includegraphics[scale=0.8]{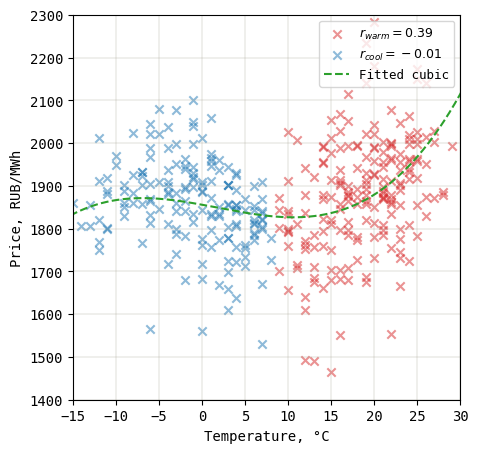}
\caption{\label{fig:sv_x_2}Price vs Temperature}
\end{figure}

We want to examine the correlation between the air temperature and consumer prices. For that we will combine the price and temperature hourly data as two features and then study their relationship. The results of the analysis are shown in the Figure \ref{fig:sv_x_2}. First, we applied the K-Means clustering algorithm \cite[p. 521]{12} to see if there are meaningful clusters in the data. We can clearly see that data has a rather V-shaped form with two distinct clusters: prices for warm and cool air temperatures. For each cluster we computed the coefficient of correlation \cite[p. 265]{25}. We can see that the consumer price \textit{does} depend on the air temperature, at least for the warm cluster:

\begin{itemize}[noitemsep]
\item High air temperatures cluster: coefficient of correlation $r_{warm} = 0.39$,
\item Low air temperatures cluster: coefficient of correlation $r_{cool} = -0.01$.
\end{itemize}

The price tends to be higher for higher and lower air temperatures, which confirms the demand expectations: electricity demand is higher during the hot and cold seasons of the year (both cooling and heating are required), while the demand is minimal during semi-seasons when minimal heating and cooling are required. At the same time, the price responds more to the higher air temperatures.

Second, for the whole data we fitted a third degree polynomial:

\begin{equation} \label{eq:9}
y_{t} \propto X_{t}^3,
\end{equation}
where $X_{t}$ is the hourly air temperature at time $t$. We tested linear, parabolic and cubic functions and chose the cubic. We did not want our new model to either under- (linear, parabolic) or overfit (degrees larger than 3). Also, observe that the domain of \ref{eq:9} is almost surely limited within a reasonable interval of air temperatures, which means that the model should not either underfit or overfit, since it is simply not defined outside of this interval. Finally, taking into account computation complexity with Bayesian inference with Stan, the cubic polynomial seems to be a proper choice.

\subsection{Weekday}

\begin{figure}
\centering
\includegraphics[scale=0.85]{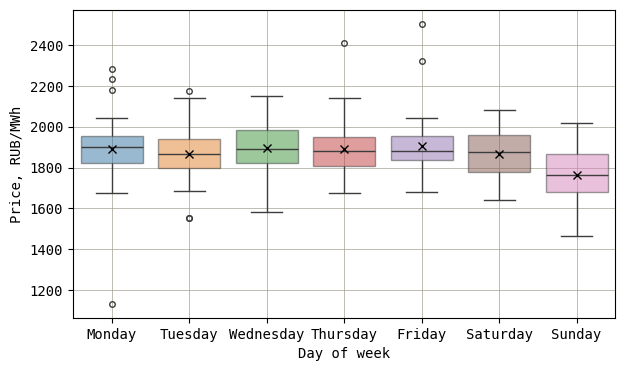}
\caption{\label{fig:sv_x_3}Correlation between the price and day of the week}
\end{figure}

Our next possible regressor might be the day of the week. It may well be expected that the electricity demand is not constant during the week with peak and off-peak days, just like it is not constant during any given day. Formally, both the consumer price and day of the week are discrete RV, and we might apply the $\chi^2$-test for their independence \cite[p. 714]{25}. However, since the price's cardinality is much larger than that of the day of the week, the $\chi^2$-test is not practically applicable, since most likely the contingency table will not satisfy the minimal count assumption of the test ($N_{Cell \neq 0} \geq 80\%$). Instead, we assess their correlation with box plots -- see Figure \ref{fig:sv_x_3}.  

The box plots confirms that the mean price tends to be lower on Sundays and the volatility tends to be lower on Fridays, which suggests that the consumer price depends on the day of the week:

\begin{equation} \label{eq:10}
y_{t} \propto D_{t},
\end{equation}
where $D_{t}$ is the day of the week at time $t$. 

\subsection{Autoregressive Component}

\begin{figure}[hb]
\centering
\includegraphics[scale=0.85]{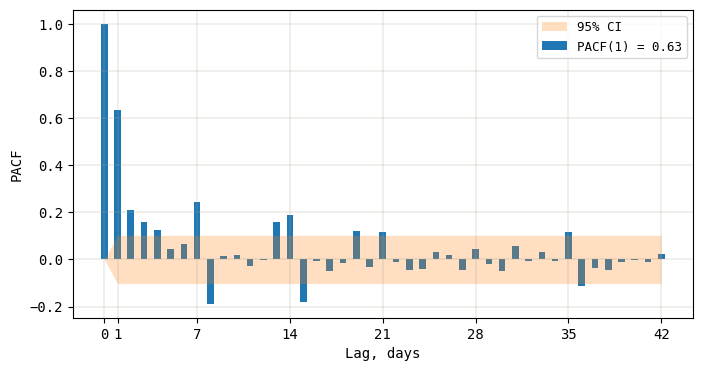}
\caption{\label{fig:sv_x_4}PACF}
\end{figure}

We will examine the correlation between the prices at time $t$ (today) and $t-1$ (yesterday). For that, we will compute the PACF \cite{6} which is the correlation between two observations that the shorter lags between those observations do not explain, i.e., the partial correlation for each lag is the unique correlation between those two observations after removing out the intervening correlations. The plot of PACF for the price zone 1 and peak hour \myhash 14 (modeling dataset) is shown in the Figure \ref{fig:sv_x_4}.

Since our price is not stationary (see \ref{subsec_test_stationarity}), the PACF decays rather slowly during the period of 42 days (6 weeks). We have shown above that the price \textit{is} correlated with the day of the week. Observe that the PACF has spikes exactly every 7 days (1 week) which further supports the correctness of possible inclusion of the day of the week as an exogenous regressor.

PACF suggests that using at least 7 or more lags (new features) might be possible. We already decided to include the day of the week to explain the seasonality of the price. The strongest correlation is seen for the lag of one ($0.63$), while other lags are much weaker correlated with the price. Taking into account computational complexity of Bayesian inference with Stan, using only the lag of one day seems to be a reasonable compromise. In other words, we will consider an autoregressive model with lag of one AR(1) as the autoregressive component.

A first-order autoregressive model AR(1) is the following \cite{4}:

\begin{equation} \label{eq:11}
y_{t} = \mathcal{N} \left ( \alpha + \beta y_{t-1}, \sigma \right ),
\end{equation}
where $\alpha$ and $\beta$ are the intercept and slope of the autoregressive component and $\sigma \sim \mathcal{N}(0, 1)$ (constant normal volatility). Since we model volatility as a stochastic process (not constant), but the AR-type models assume homoscedastic volatility \cite[p. 1055]{8}, we will consider only the mean value of the AR(1) model.

\subsection{Modeling}

We have shown that the consumer price is correlated with 1) air temperature, 2) day of the week, and 3) with itself with at least lag of one day. We propose a new model SV X, which extends our SV Baseline model \ref{eq:6} with the three exogenous regressors \ref{eq:9}, \ref{eq:10}, \ref{eq:11}:

\begin{equation} \label{eq:12}
y_{t} \sim \mathcal{N} \left ( \bar{y} + \alpha y_{t-1} + \beta_3 X_{t-1}^3 + \beta_2 X_{t-1}^2 + \beta_1 X_{t-1} + \gamma D_{t} + \xi, e^{h_{t} / 2} \right ),
\end{equation}
where $X_{t-1}$ is the hourly air temperature at time $t-1$. To prevent target leakage, the temperature readings are lagged one day behind ($t-1$); $D_{t}$ is the day of the week at time $t$.

Thus, we are introducing 6 new model parameters for 3 exogenous regressors:
\begin{itemize}[noitemsep]
\item $\alpha$, autoregressive component;
\item $\beta_{i = 1 \ldots 3}$, air temperature regressor. The unit of air temperature is °C;
\item $\gamma$, day of the week regressor. The weekdays are numbered from 0 (Monday) to 6 (Sunday);
\item $\xi$, constant term (intercept) for all exogenous regressors.
\end{itemize}

Recall that we assumed the constant mean $\bar{y}$ in the SV Baseline model which is not correct for our non-stationary price. SV X takes care of this drawback by "mixing" exogenous regressors to the constant price mean in a regression-like manner.

\begin{figure}[hb]
\centering
\includegraphics[scale=0.65]{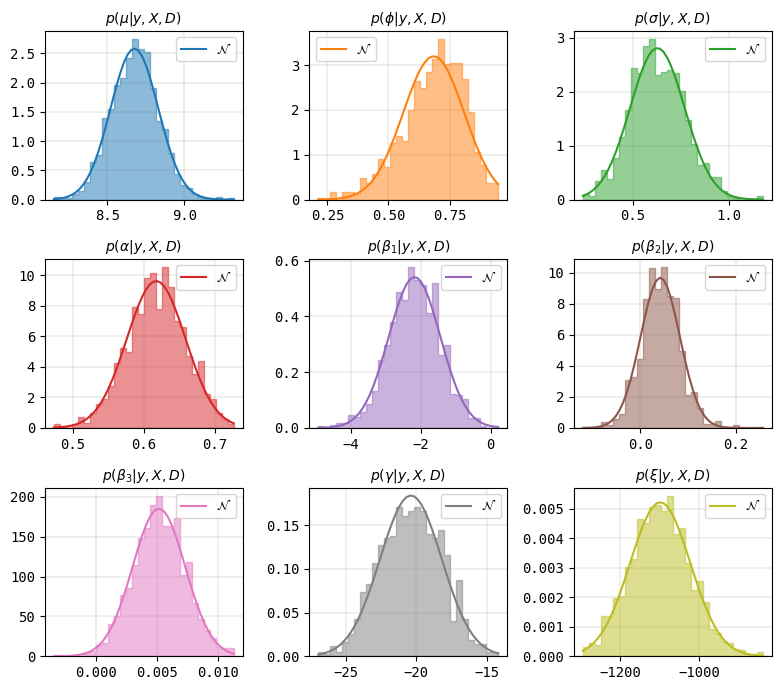}
\caption{\label{fig:sv_x_5}SV X: Posterior distributions of estimated parameters}
\end{figure}

After running the Stan code, we have obtained the posterior distributions of all SV X model's parameters -- see Figure \ref{fig:sv_x_5}.

\begin{figure}[hb]
\centering
\includegraphics[scale=0.8]{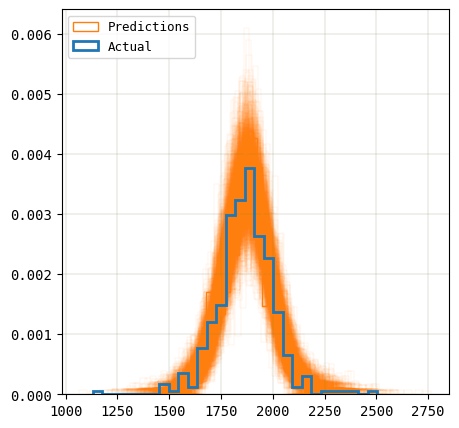}
\caption{\label{fig:sv_x_12}SV X: 1,000 predicted and actual price distributions}
\end{figure}

We can now use the estimated model's parameters to compute PDD and generate in-sample predictions for the modeling time frame. 1,000 density plots sampled from the PDD of the consumer price are shown in the Figure \ref{fig:sv_x_12}.

\begin{figure}[hb]
\centering
\includegraphics[scale=0.72]{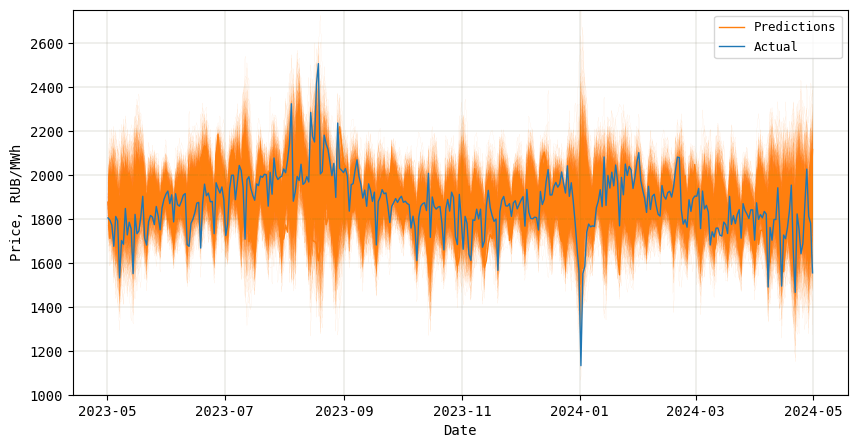}
\caption{\label{fig:sv_x_6}SV X: Actual price and 1,000 predictions}
\end{figure}

Observe, that the actual distribution (density) of the consumer prices is well within the predicted density plots, which confirms the correctness of modeling. 1,000 trace plots sampled from the PDD of the consumer prices are shown in the Figure \ref{fig:sv_x_6}.

\begin{figure}[hb]
\centering
\includegraphics[scale=0.75]{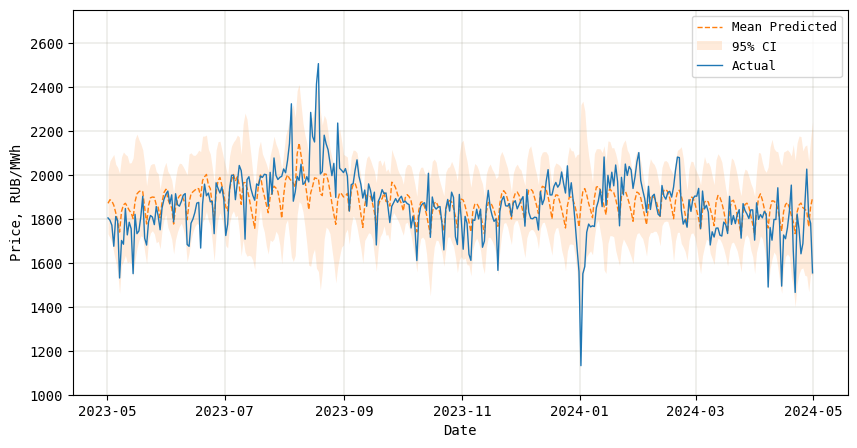}
\caption{\label{fig:sv_x_7}SV X: 95\% CI for 1,000 predictions}
\end{figure}

The 95\% CI for all 1,000 predictions is shown in the Figure \ref{fig:sv_x_7}.

\begin{figure}[hb]
\centering
\includegraphics[scale=0.70]{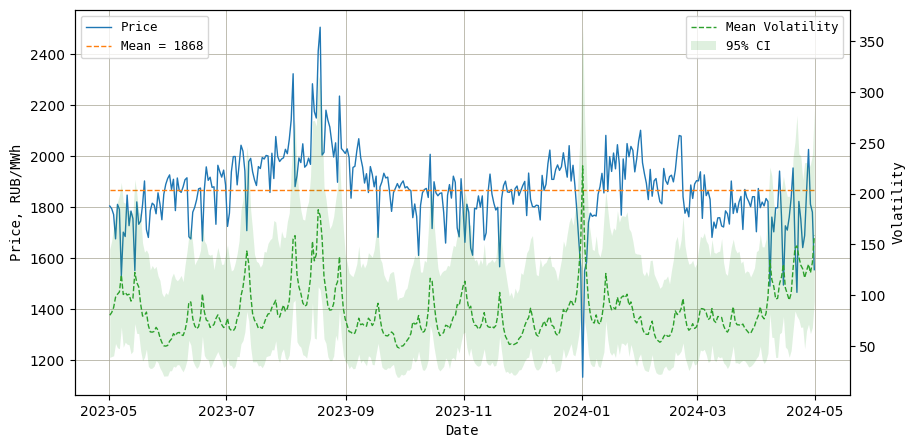}
\caption{\label{fig:sv_x_8}SV X: Consumer prices and learned volatility}
\end{figure}

The model has learned the latent stochastic volatility process quite correctly. The learned volatility $e^{h_{t} / 2}$ (both mean and 95\% CI) is shown in the Figure \ref{fig:sv_x_8}.

\begin{figure}
\centering
\includegraphics[scale=0.8]{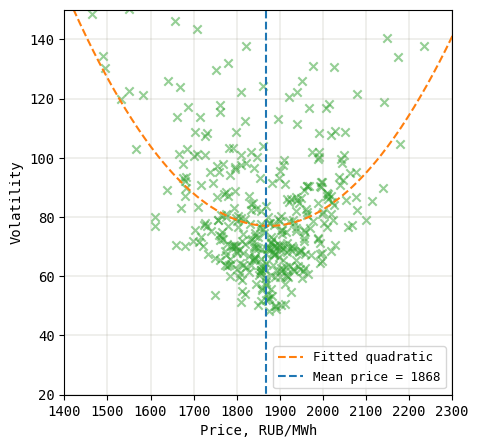}
\caption{\label{fig:sv_x_9}SV X: Learned volatility vs Consumer price}
\end{figure}

To understand the behavior of the volatility we will visually check the correlation between the price and volatility -- see Figure \ref{fig:sv_x_9}.

We can clearly see that volatility \textit{does} depend on consumer price, having a rather V-shaped dependency: the volatility tends to increase for the prices higher and lower than the mean price, while it's minimal for the prices around the mean.

\subsection{Goodness-of-fit}

The metrics for the in-sample predictions made by the SV X model are shown in the Table \ref{tab:sv_x_1}. SV Baseline model's metrics are shown for comparison.

\begin{longtable}{l||r|r}
\hline
Metric & SV Baseline & SV X \\
\hline \hline
MAE & 99.18 & \textbf{85.23} (-14.06\%) \\
RMSE & 137.04 & \textbf{117.91} (-13.96\%) \\
\hline
\caption{\label{tab:sv_x_1}SV X: MAE and RMSE for predicted consumer prices}
\end{longtable}

\begin{figure}
\centering
\includegraphics[scale=0.8]{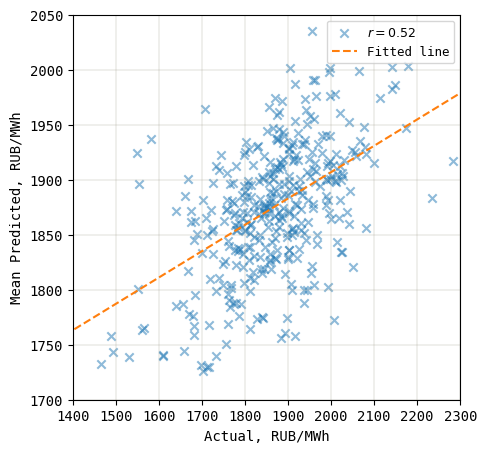}
\caption{\label{fig:sv_x_10}SV X: Predicted vs Actual price}
\end{figure}

We also want to check the correlation between the predicted mean price and actual price -- see Figure \ref{fig:sv_x_10}. Observe, that the correlation is much stronger, as compared to the SV Baseline model.

\begin{figure}[hb]
\centering
\includegraphics[scale=0.55]{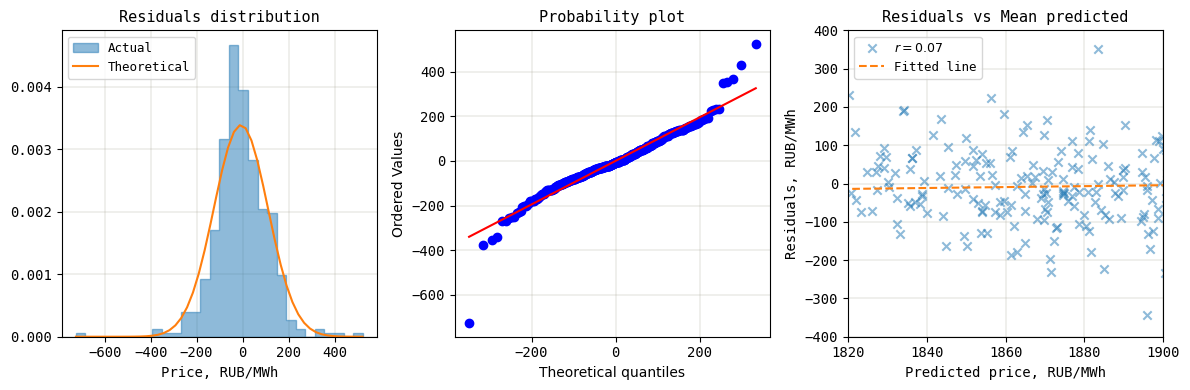}
\caption{\label{fig:sv_x_11}SV X: Residuals plots}
\end{figure}

We should also check the distribution of residuals, probability plot, and correlation between residuals and mean predicted prices -- see Figure \ref{fig:sv_x_11}. Residuals look to be distributed more or less normally with several outliers. Also, residuals are almost not correlated with predictions (homoscedastic).

\subsection{Model Interpretation}

Finally, we will look at PD of the price from individual exogenous regressors. The idea of such model interpretation is to predict the price for a number of fixed values for each regressor (within its domain), i.e. to compute the predictions as a function of some feature, keeping other (complementary) features intact \cite[8.1 Partial Dependence Plot (PDP)]{28}. If the prediction function responds to different values of a feature, then this feature is considered as important, otherwise not. These predictions are ICEs for each feature of interest. PD is simply the average of ICEs for each feature of interest over all predictions (number of observations). The main advantage of this model interpretation method is that we do not make any assumptions about linear / non-linear dependencies of the price from the features -- we simply assess this dependence ad hoc from real model predictions.

PD method has an assumption of feature independence between each other. The coefficient of correlation between the Air temperature and Day of the week is $-0.024$, which means that at least there is almost no linear dependence between these exogenous factors.

\begin{figure}[hb]
\centering
\includegraphics[scale=0.8]{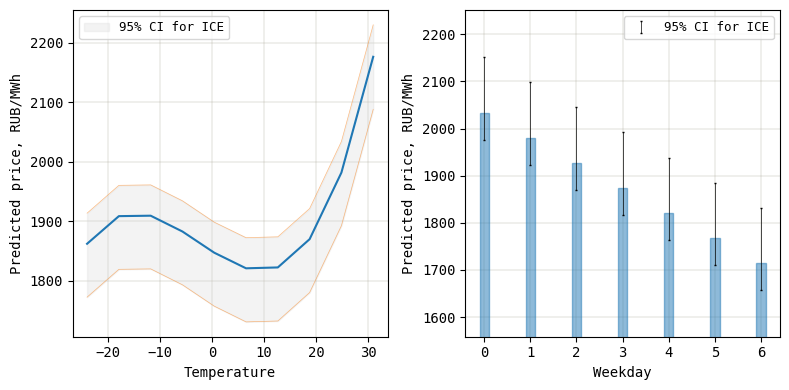}
\caption{\label{fig:sv_x_13}SV X: PD Plots}
\end{figure}

The PD plots are shown in the Figure \ref{fig:sv_x_13}. The plots confirm that the SV X model's predictions respond to the exogenous regressors Air temperature and Day of the week.

We will \textit{accept} the SV X model as a \textit{better} fit for our price data with exogenous regressors, as compared to the SV Baseline model.

\newpage

\section{CROSS-VALIDATION}

We have trained and tested our models on the same fixed year-long modeling time frame. There is a common practice in ML to train and test models on different portions of the whole data to generate the distribution of metric(s) to ensure model robustness. This process is known as cross-validation \cite[p. 202]{12}: all data is divided into a number of non-intersecting subsets of samples (folds) some of which are used for training and the rest for testing. For data with i.i.d. samples, i.e., when the samples are independent and thus the order of samples is not important, the order of folds is also not important. This is not the case for time series data, where training folds must not precede the testing folds, otherwise the target leakage will occur and the results cannot be trusted.

\subsection{Strategy} \label{subsec_strategy}

Model combinations to cross-validate are shown in the Table \ref{tab:cross_val_1}.

\begin{longtable}{c||l|l|l}
\hline
Combination \myhash & Model family & Hour & Price zone \\
\hline \hline
1 & \multirow{4}{*}{SV Baseline} & Peak & 1 (European) \\
2 & & Peak & 2 (Siberian) \\
3 & & Off-peak & 1 \\
4 & & Off-peak & 2 \\
\hline
5 & \multirow{4}{*}{SV X} & Peak & 1 \\
6 & & Peak & 2 \\
7 & & Off-peak & 1 \\
8 & & Off-peak & 2 \\
\hline
\caption{\label{tab:cross_val_1}Model combinations to cross-validate}
\end{longtable}

Time frame: 23.06.2014 -- 30.04.2024 (3600 days), approx. ten years back from now. The number of date sliding windows, and thus the number of train-test folds, can be computed as $N_{w} = \frac{3600 - 30 \times 12}{30 \times 3} = 36$. All 36 train-test fold time frames, as well as the consumer price and air temperature data descriptive statistics for both price zones over the cross-validation time frame can be found in the author's computations notebook \cite{10}. The cross-validation algorithm \ref{alg:1} is shown below.

\begin{algorithm}[h]
\begin{spacing}{1.5}
\caption{Cross-validation} \label{alg:1}
\begin{algorithmic}[1]
\State Split the data into train-test folds.
\ForAll{Model $\in [1, 8]$} \Comment{4 SV Baseline + 4 SV X models}
\ForAll{Fold $\in [1, N_{w}]$} \Comment{$N_{w} = 36$}
\begin{itemize}[noitemsep]
\item Fit on train fold (360 days -- approx. 1 year); \Comment{80/20 split}
\item Predict on test fold (90 days -- approx. 3 months = 1 quarter) using the most recent log volatility learned during training (90 points);
\item Compute metrics: actual (test fold) vs mean predicted.
\end{itemize}
\EndFor
\EndFor
\end{algorithmic}
\end{spacing}
\end{algorithm}

Hour and temperature hyperparameters used for cross-validation are shown in the Table \ref{tab:cross_val_2}.

\begin{longtable}{l||r|r|l}
\hline
Price zone & Peak hour & Off-peak hour & Air temperature city \\
\hline \hline
1 (European) & \myhash 11 & \myhash 3 & Moscow \\
2 (Siberian) & \myhash 16 & \myhash 1 & Novosibirsk \\
\hline
\caption{\label{tab:cross_val_2}Hyperparameters used for cross-validation}
\end{longtable}

Observe that the peak hour for the price zone 1 differs from the one used in modeling (\myhash 14). This can be explained by a different (10 times wider) time frame used for cross-validation as compared to the time frame used for modeling.

\subsection{SV Baseline}

During cross-validation, for each SV Baseline model and for each of the 36 folds, posterior distributions of all model parameters are obtained (on train fold) and then the PDD for predictions is computed from which 1,000 samples is drawn (on test fold). The MAE and RMSE are computed, as in modeling, using the averaged samples of the predicted consumer prices for time $t$ -- see Table \ref{tab:cross_val_3}.

\begin{longtable}{c||l|l|r|r}
\hline
Model family & Hour & Price zone & MAE & RMSE \\
\hline \hline
\multirow{4}{*}{SV Baseline} & Off-peak hour & 1 & \textbf{104.42} & 137.50 \\
& Peak hour & 1 & 105.39 & \textbf{130.53} \\
& Peak hour & 2 & 119.14 & 146.45 \\
& Off-peak hour & 2 & 134.86 & 168.31 \\
\hline
\caption{\label{tab:cross_val_3}SV Baseline: MAE and RMSE results for cross-validation}
\end{longtable}

Both metrics show that on average the SV Baseline models for Price zone 1 (European) have higher quality than for Price zone 2 (Siberian).

\subsection{SV X}

Same as with SV Baseline, the MAE and RMSE are computed using the averaged samples of the predicted consumer prices for time $t$ -- see  Table \ref{tab:cross_val_4}.

\begin{longtable}{c||l|l|r|r}
\hline
Model family & Hour & Price zone & MAE & RMSE \\
\hline \hline
\multirow{4}{*}{SV X} & Off-peak & 1 & \textbf{93.92} & 129.59 \\
& Peak & 1 & 99.45 & \textbf{123.59} \\
& Peak & 2 & 115.56 & 142.10 \\
& Off-peak & 2 & 118.18 & 150.73 \\
\hline
\caption{\label{tab:cross_val_4}SV X: MAE and RMSE results for cross-validation}
\end{longtable}

As with the SV Baseline model, both metrics for the SV X models show that on average the SV X models for Price zone 1 (European) have higher quality than for Price zone 2 (Siberian).

\subsection{Model Comparison}

\begin{figure}[h]
\centering
\includegraphics[scale=0.9]{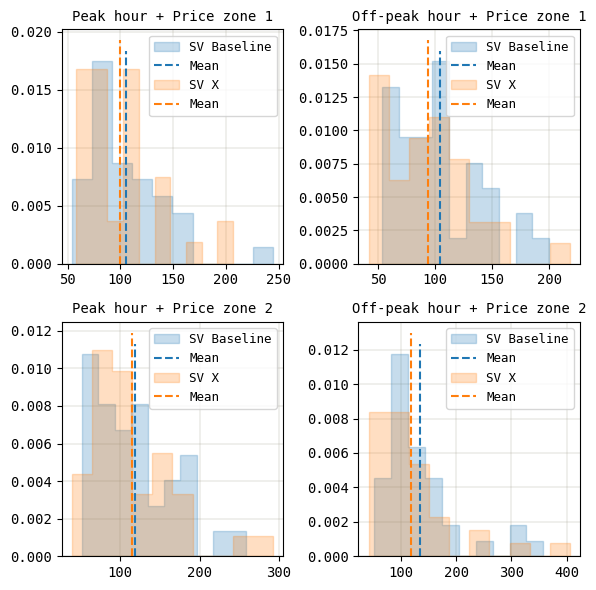}
\caption{\label{fig:cross_val_1}MAE: Cross-validation for all models}
\end{figure}

\begin{figure}[h]
\centering
\includegraphics[scale=0.9]{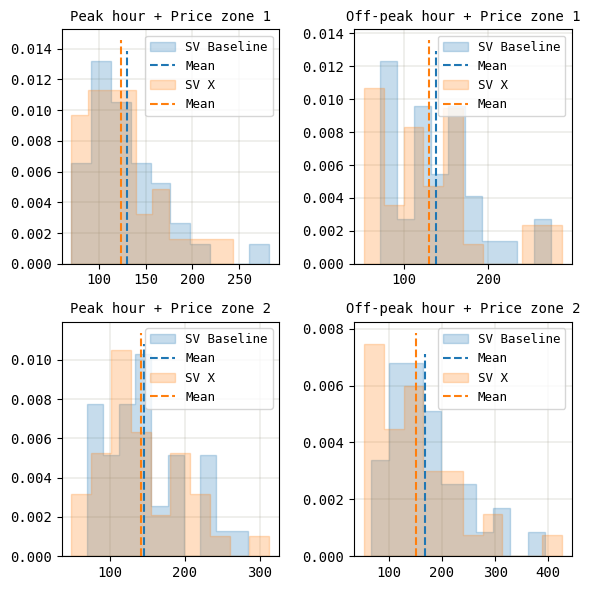}
\caption{\label{fig:cross_val_2}RMSE: Cross-validation for all models}
\end{figure}

We have generated metrics distributions during models cross-validation -- see Figure \ref{fig:cross_val_1} for MAE and Figure \ref{fig:cross_val_2} for RMSE. Each distribution contains 36 values, as per the number of cross-validation date sliding windows.

We would also like to check the statistical significance of the difference in the metrics. We will use a non-parametric one-tailed Mann-Whitney U test for two independent samples \cite[p. 758]{25} to see if the distributions of metrics for SV Baseline and SV X models are identical or have a shift between each other.

$\alpha = 0.05$

$\displaystyle \mathcal{H}_0: F_{X_1}(x) =  F_{X_2}(x)$ (distributed identically)

$\displaystyle \mathcal{H}_1: F_{X_1}(x) =  F_{X_2}(x + \Delta x)$ (distributed with positive shift $\Delta x$), where

$F_{X_1}(x)$ is the distribution of a given metric for SV Baseline model; $F_{X_2}(x)$ is the distribution of a given metric for SV X model.

Statistic: Mann-Whitney's $U$

Null distribution: no assumption; generated empirically by permutation test; for large samples it can be approximated by $\mathcal N \left (\mu = \frac{n_1 n_2}{2}, \sigma^2 = \frac{n_1 n_2 (n_1 + n_2 + 1)}{12} \right )$.

The results of the MWU tests are shown in the Table \ref{tab:cross_val_5}.

\begin{longtable}{l||r|r|l}
\hline
Metric & Statistic & $p$-value & Result at $\alpha$ \\
\hline \hline
MAE & 11785 & 0.0225 & $\Delta x > 0$ \\
RMSE & 11575 & 0.0439 & $\Delta x > 0$ \\
\hline
\caption{\label{tab:cross_val_5}MWU tests results}
\end{longtable}

We have rejected the null hypothesis at the significance level of 0.05 in favor of the alternative hypothesis -- SV Baseline's metrics are shifted to the right relative to SV X's metrics (= worse).

The summary of metrics for all models computed during cross-validation is shown in the Table \ref{tab:cross_val_6}.

\begin{longtable}{l||l|l|r|r}
\hline
Metric & Hour & Price zone & SV Baseline & SV X \\
\hline \hline
\multirow{4}{*}{MAE} & Peak & 1 & 105.39 & \textbf{99.45} \\
& Peak & 2 & 119.14 & \textbf{115.56} \\
& Off-peak & 1 & 104.42 & \textbf{93.92} \\
& Off-peak & 2 & 134.86 & \textbf{118.18} \\
\hline
\multicolumn{3}{r|}{Average} & 115.95 & \textbf{106.78} (-7.91\%) \\
\hline \hline
\multirow{4}{*}{RMSE} & Peak & 1 & 130.53 & \textbf{123.59} \\
& Peak & 2 & 146.45 & \textbf{142.10} \\
& Off-peak & 1 & 137.50 & \textbf{129.59} \\
& Off-peak & 2 & 168.31 & \textbf{150.73} \\
\hline
\multicolumn{3}{r|}{Average} & 145.69 & \textbf{136.50} (-6.31\%) \\
\hline
\caption{\label{tab:cross_val_6}Cross-validation: Summary of metrics for all models}
\end{longtable}

The results of cross-validation:
\begin{enumerate}[noitemsep]
\item SV X's MAE is 7.91\% less (= better) than SV Baseline's MAE on average;
\item SV X's RMSE is 6.31\% less (= better) than SV Baseline's RMSE on average;
\item These differences are statistically significant at $\alpha = 0.05$;
\item Overall, the SV X model family is a \textit{better} fit for our price data with the air temperature, day of the week, and lagged price as the exogenous regressors, as compared to the SV Baseline model family without exogenous regressors.
\end{enumerate}

\newpage

\section{FORECASTING}

Having built and cross-validated our models, we would finally like to generate forecasts for the nearest day(s) ahead with them.

\subsection{Strategy}

Just like with cross-validation, we will generate forecasts for 8 model combinations (see Table \ref{tab:cross_val_1}). First train time frame: 01.05.2023 -- 30.04.2024 (same as used for modeling). Forecast time frame: 01.05.2024 -- 07.05.2024 (1 week ahead). The forecasting algorithm \ref{alg:2} is shown below.

\begin{algorithm}
\begin{spacing}{1.5}
\caption{Forecasting} \label{alg:2}
\begin{algorithmic}[1]
\State Set the train time frame to the first train time frame.
\ForAll{Model $\in [1, 8]$} \Comment{4 SV Baseline + 4 SV X models}
\ForAll{Day $\in [1, 7]$} \Comment{7 days = 1 week}
\begin{itemize}[noitemsep]
\item Fit the model on the train time frame (1 year);
\item Generate \textit{one} day-ahead prediction (forecast) using the most recent log volatility learned during training (1 point), and populate the forecasted prices;
\item Move the train time frame forward by \textit{one} day.
\end{itemize}
\EndFor
\item Compute metrics: actual vs mean forecasted;
\EndFor
\end{algorithmic}
\end{spacing}
\end{algorithm}

\subsection{Results}

\begin{figure}
\centering
\includegraphics[scale=0.75]{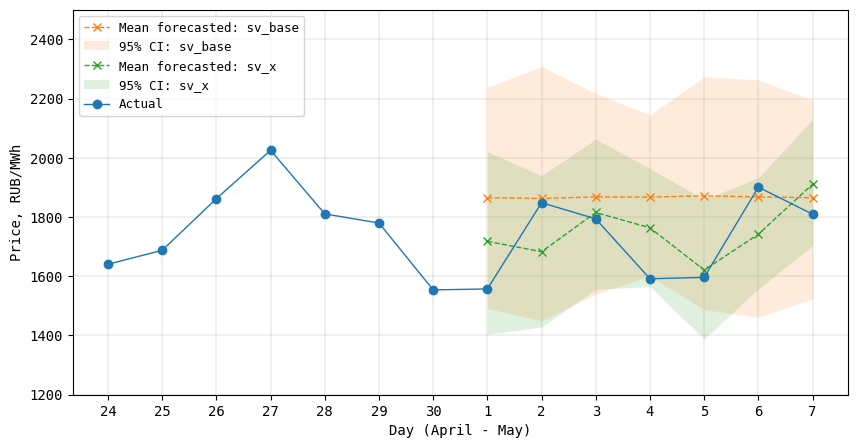}
\caption{\label{fig:forecast_1}Forecasts: Peak hour + Price zone 1}
\end{figure}

\begin{figure}
\centering
\includegraphics[scale=0.75]{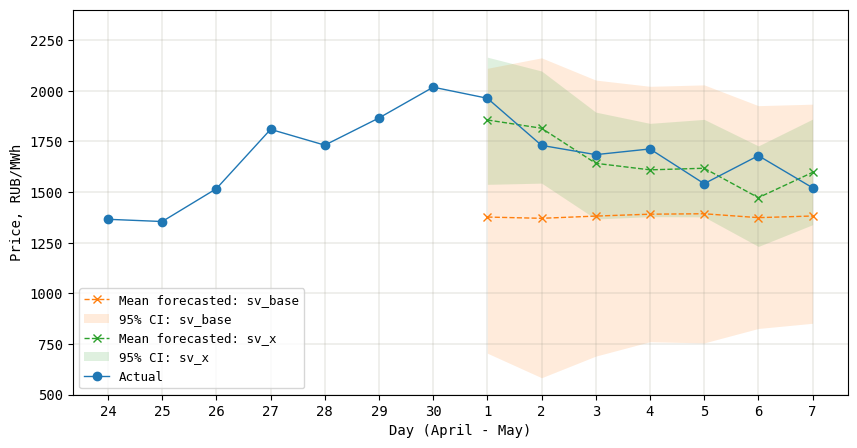}
\caption{\label{fig:forecast_2}Forecasts: Peak hour + Price zone 2}
\end{figure}

\begin{figure}
\centering
\includegraphics[scale=0.75]{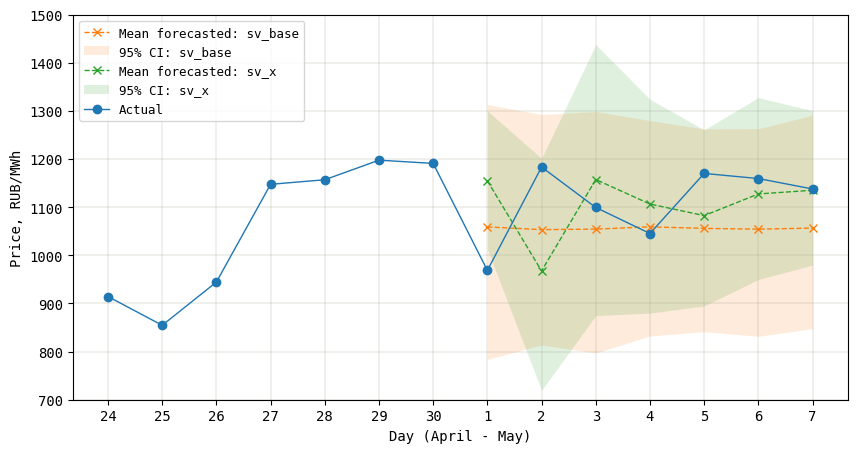}
\caption{\label{fig:forecast_3}Forecasts: Off-peak hour + Price zone 1}
\end{figure}

\begin{figure}
\centering
\includegraphics[scale=0.75]{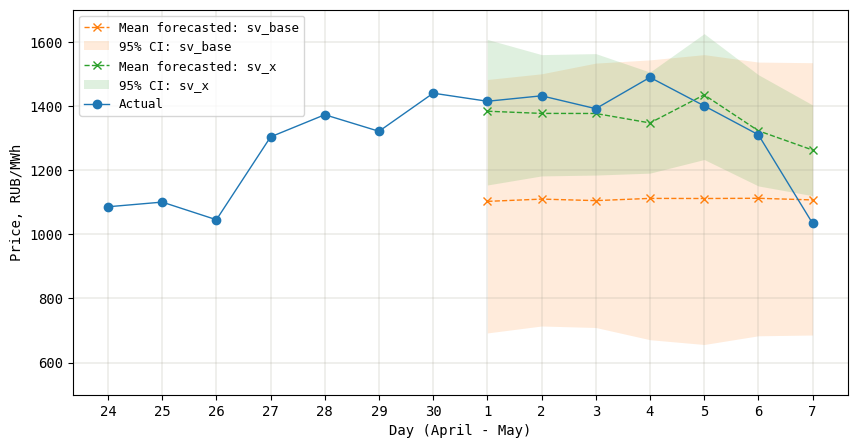}
\caption{\label{fig:forecast_4}Forecasts: Off-peak hour + Price zone 2}
\end{figure}

For all model combinations and for each day ahead we have drawn 1,000 forecasts from the day's PDD. The plots of mean of each 1,000 draws and 95\% CI for forecasts are shown in the Figures \ref{fig:forecast_1}, \ref{fig:forecast_2}, \ref{fig:forecast_3}, \ref{fig:forecast_4}.

For computing the metrics, all draws were averaged for each day (same as in modeling and cross-validation). The summary of metrics is shown in the Table \ref{tab:forecast_1}.

\begin{longtable}{l||l|l|r|r}
\hline
Metric & Hour & Price zone & SV Baseline & SV X \\
\hline \hline
\multirow{4}{*}{MAE} & Peak & 1 & 148.06 & \textbf{114.95} \\
& Peak & 2 & 309.28 & \textbf{99.56} \\
& Off-peak & 1 & \textbf{82.84} & 91.94 \\
& Off-peak & 2 & 265.35 & \textbf{73.97} \\
\hline
\multicolumn{3}{r|}{Average} & 201.38 & \textbf{95.11} (-52.77\%) \\
\hline					
\multirow{4}{*}{RMSE} & Peak & 1 & 191.49 & \textbf{130.60} \\
& Peak & 2 & 338.86 & \textbf{110.59} \\
& Off-peak & 1 & \textbf{91.02} & 117.76 \\
& Off-peak & 2 & 281.16 & \textbf{105.44} \\
\hline
\multicolumn{3}{r|}{Average} & 225.63 & \textbf{116.10} (-48.54\%) \\
\hline
\caption{\label{tab:forecast_1}Forecasting: Summary of metrics for all models}
\end{longtable}

The results of forecasting:
\begin{enumerate}[noitemsep]
\item SV X's MAE is 52.77\% less (= better) than SV Baseline's MAE on average;
\item SV X's RMSE is 48.54\% less (= better) than SV Baseline's RMSE on average;
\item The SV X model is a \textit{better} fit for Peak hour + Price zone 1, Peak hour + Price zone 2, Off-peak + Price zone 2.
\item The SV X model is a \textit{worse} fit for Off-peak hour + Price zone 1.
\end{enumerate}

In general, all models are able to learn the price change -- 95\% CI for all sampled forecasts cover the actual price. SV X model family provides a narrower forecast CI as compared to SV Baseline model family. The reason for this is that for the SV X model the latent log volatility $h_{t}$ is learned w.r.t. the mean price being not simply a constant value but a function of the mean and exogenous regressors, as opposed to the SV Baseline model. As a result, $h_{t}$ can "follow" the price change more precisely.

\include{sec_conclusion}
\newpage

\section{CONCLUSION AND FUTURE WORK}

We have built and tested our models for two market hours: peak and off-peak, and two price zones: 1 (European), 2 (Siberian). The results are a good estimation and understanding of the models' behavior. On average, the family of SV X models is a \textit{better} fit for predicting the price over time with exogenous regressors (see Tables \ref{tab:cross_val_6} and \ref{tab:forecast_1}). The forecasting results show that for the Price zone 1 + Off-peak hour the better performance is provided by the SV Baseline model, while for the rest three scenarios the best performer is the SV X model. Thus, we should build similar models for all 24 hours for the day-ahead for both price zones and then validate and compare all obtained models to choose the best performer for future forecasting. Ideally, we should build 48 hourly models tailored to each price zone and each hour of the day.

The cross-validation and forecasting results confirm the applicability and robustness of the enhanced SV X model. This model may be used in financial derivative instruments for hedging the risk associated with electricity trading.

The author is developing their own ML library in C++ called EasyML \cite{9}. The main motivation is to build a compact, flexible and fast library tailored for subject matter tasks. The library already covers the main linear models: Linear Regression, Logistic Regression, AR(p). There are also data transformers already available: Standard Scaler ($z$-scores) and Time Series Feature Extractor (lags).

We have introduced only one weather regressor -- air temperature. At the same time, it might well be expected that other climate and weather factors drive the electricity demand: humidity, precipitation, solar irradiance, wind speed, etc. The future research might include these factors as new regressors into modeling. As the number of regressors increase, the computational complexity of training and generating predictions increases too. We have studied 8 models, while in fact their number might be up to 48 (96 including SV Baseline). The need to re-train this number of models each and every day to generate the next day-ahead forecast makes the author believe that a task-tailored library will increase computation speed thus allowing to build more complex models and validate them more rigorously and promptly in production-like scenarios.

\newpage

\phantomsection
\addcontentsline{toc}{section}{REFERENCES}
\bibliographystyle{IEEEtran}
\bibliography{main}

\newpage

\phantomsection
\addcontentsline{toc}{section}{APPENDIX}
\section*{APPENDIX}

\addcontentsline{toc}{subsection}{A.1. Stan Code for SV Baseline Model}
\subsection*{A.1. Stan Code for SV Baseline Model}

\begin{lstlisting}[language={}, caption=SV Baseline: Fit method, label={lst:1}]
/**
 * @file sv_base_fit.stan
 * @author Andrei Batyrov (arbatyrov@edu.hse.ru)
 * @brief SV Baseline model: Fit method
 * Exogenous regressors: None
 * @version 0.1
 * @date 2024-05-08
 * 
 * @copyright Copyright (c) 2024
 * 
 */

data
{
  int<lower=1> N; // Number of train time points (equally spaced)
  vector[N] y;    // Vector of train prices at time t
  real y_mean;    // Mean price
}

parameters
{
  real mu;                     // Mean log volatility
  real<lower=-1, upper=1> phi; // Persistence of volatility
  real<lower=0> sigma;         // White noise shock scale
  vector[N] h_std;             // Standardized log volatility at time t
}

transformed parameters
{
  // Log volatility at time t; now h ~ normal(0, sigma)
  vector[N] h = h_std * sigma;
  h[1] = h[1] / sqrt(1 - pow(phi, 2)); // Rescale h[1]
  h = h + mu;
  for (t in 2:N)
  {
    h[t] = h[t] + phi * (h[t - 1] - mu);
  }
}

model
{
  // Priors recommended by Stan
  phi ~ uniform(-1, 1);
  sigma ~ cauchy(0, 5);
  mu ~ cauchy(0, 10);
  h_std ~ std_normal();
  
  // Model
  y ~ normal(y_mean, exp(h / 2));
}
\end{lstlisting}

\begin{lstlisting}[language={}, caption=SV Baseline: Predict method, label={lst:2}]
/**
 * @file sv_base_predict.stan
 * @author Andrei Batyrov (arbatyrov@edu.hse.ru)
 * @brief SV Baseline model: Predict method
 * Exogenous regressors: None
 * @version 0.1
 * @date 2024-05-08
 * 
 * @copyright Copyright (c) 2024
 * 
 */

data
{
  int<lower=1> N_pred; // Number of test time points (equally spaced)
  vector[N_pred] h;    // Vector of learned log volatility at time t
  real y_mean;         // Learned mean price
}

generated quantities
{
  vector[N_pred] y_pred;

  // Generate posterior predictive distribution
  for (t in 1:N_pred)
  {
      y_pred[t] = normal_rng(y_mean, exp(h[t] / 2));
  }
  
}
\end{lstlisting}

\addcontentsline{toc}{subsection}{A.2. Stan Code for SV X Model}
\subsection*{A.2. Stan Code for SV X Model}

\begin{lstlisting}[language={}, caption=SV X: Fit method, label={lst:3}]
/**
 * @file sv_x_fit.stan
 * @author Andrei Batyrov (arbatyrov@edu.hse.ru)
 * @brief SV Exogenous model: Fit method
 * Exogenous regressors:
 * - Temperature
 * - Weekday
 * - AR(1)
 * @version 0.1
 * @date 2024-05-08
 * 
 * @copyright Copyright (c) 2024
 * 
 */

data
{
  int<lower=1> N;        // Number of train time points (equally spaced)
  vector[N] y;           // Vector of train prices at time t
  real y_mean;           // Mean price
  vector[N] Temperature; // Vector of train Temperature at time t
  vector[N] Weekday;     // Vector of train Weekday at time t
}

parameters
{
  real mu;                     // Mean log volatility
  real<lower=-1, upper=1> phi; // Persistence of volatility
  real<lower=0> sigma;         // White noise shock scale
  vector[N] h_std;             // Standardized log volatility at time t
  real alpha;                  // Param of AR(1) component
  real beta_1;                 // Params of Temperature regressor
  real beta_2;
  real beta_3;
  real gamma;                  // Param of Weekday regressor
  real xi;                     // Intercept for all exogenous regressors
  
}

transformed parameters
{
  // Log volatility at time t; now h ~ normal(0, sigma)
  vector[N] h = h_std * sigma;
  h[1] = h[1] / sqrt(1 - pow(phi, 2)); // Rescale h[1]
  h = h + mu;
  for (t in 2:N)
  {
    h[t] = h[t] + phi * (h[t - 1] - mu);
  }
}

model
{
  // Priors recommended by Stan
  phi ~ uniform(-1, 1);
  sigma ~ cauchy(0, 5);
  mu ~ cauchy(0, 10);
  h_std ~ std_normal();
  
  // Model
  // We do not model y[t = 1],
  // since we do not have y[t - 1 = 0] and Temperature[t - 1 = 0]
  // So assume the first value without Temperature and AR(1)
  // and continue from index 2
  y[1] ~ normal(y_mean + gamma * Weekday[1], 
                exp(h[1] / 2));
  for (t in 2:N)
  {
    y[t] ~ normal(y_mean + 
                  alpha * y[t - 1] +
                  beta_3 * pow(Temperature[t - 1], 3) + 
                  beta_2 * pow(Temperature[t - 1], 2) + 
                  beta_1 * Temperature[t - 1] + 
                  gamma * Weekday[t] +
                  xi, 
                  exp(h[t] / 2));
  }
}
\end{lstlisting}

\begin{lstlisting}[language={}, caption=SV X: Predict method, label={lst:4}]
/**
 * @file sv_x_predict.stan
 * @author Andrei Batyrov (arbatyrov@edu.hse.ru)
 * @brief SV Exogenous model: Predict method
 * Exogenous regressors:
 * - Temperature
 * - Weekday
 * - AR(1)
 * @version 0.1
 * @date 2024-05-08
 * 
 * @copyright Copyright (c) 2024
 * 
 */

data
{
  int<lower=1> N_pred; // Number of test time points (equally spaced)
  vector[N_pred] h;    // Vector of learned log volatility at time t
  real y_mean;         // Learned mean price
  real alpha;          // Learned params of exogenous regressors
  real beta_1;
  real beta_2;
  real beta_3;
  real gamma;
  real xi;
  vector[N_pred] Temperature_pred; // Vector of test Temperature at time t
  vector[N_pred] Weekday_pred;     // Vector of test Weekday at time t
}

generated quantities
{
  vector[N_pred] y_pred;
   
  // Generate posterior predictive distribution
  // We do not predict y_pred[t = 1],
  // since we do not have y_pred[t - 1 = 0] and Temperature[t - 1 = 0]
  // So predict the first value without Temperature and AR(1)
  // and continue from index 2
  y_pred[1] = normal_rng(y_mean + gamma * Weekday_pred[1], 
                         exp(h[1] / 2));
  for (t in 2:N_pred)
  {
    y_pred[t] = normal_rng(y_mean + 
                           alpha * y_pred[t - 1] +
                           beta_3 * pow(Temperature_pred[t - 1], 3) + 
                           beta_2 * pow(Temperature_pred[t - 1], 2) + 
                           beta_1 * Temperature_pred[t - 1] + 
                           gamma * Weekday_pred[t] +
                           xi, 
                           exp(h[t] / 2));
  }
}
\end{lstlisting}

\end{document}